\documentclass[12pt]{iopart}
\usepackage{graphicx}
\begin{document}
\title[Systematic study of the system size dependence.......]{\bf Systematic study of the system size dependence of global stopping: Role of momentum dependent interactions and symmetry energy}
\author{Sanjeev Kumar and Suneel Kumar}
\address{School of Physics and Material Science, Thapar University, Patiala-147004, Punjab (India)}
\ead{suneel.kumar@thapar.edu}
\begin{abstract}
Using the isospin-dependent quantum molecular dynamical
(IQMD) model, we systematically study the role of momentum dependent interactions in
global stopping and analyze the effect of symmetry energy in the presence of
momentum dependent interactions. For this, we simulate the reactions by varying the total mass of the system
from 80 to 394 at
different beam energies from 30 to 1000
MeV/nucleon over central and semi-central geometries. The study is carried in the presence of momentum dependent
interactions and symmetry energy by taking into account hard equation of state. The nuclear stopping is found to be sensitive towards the momentum dependent interactions and
symmetry energy at low incident energies. The momentum dependent interactions are
found to weaken the finite size effects in nuclear stopping.
\end{abstract}
%\submitto{\CPL}
\maketitle
%%%%%%%%%%%%%%%%%%%%%%%%%%%%%%%%%%%%%%%%%%%%%%%%%%%%%%%%%%%%%%%%%%%%%%%%%%%%%%%%%%%%%%%%%%%%%%%%%%%%%%%%
\section{Introduction}
The heavy-ion collisions at intermediate energies have witnessed several rare
phenomena such as multifragmentation, disappearance of flow, partial(or complete)
stopping as well as sub threshold particle production \cite{Puri94,Goss97}.
The recent advances in the radioactive nuclear beam (RNB) physics is providing scientific
community a unique opportunity to investigate
the isospin effects in heavy-ion collisions (HIC's) \cite{Li98} with respect to the above
rare phenomena \cite{Puri94,Goss97}.
Beside the many existing radioactive beam facilities, many more are being constructed or under planning, including the Cooling Storage Ring (CSR) facility at
HIRFL in China, the Radioactive Ion Beam (RNB) factory at RIKEN in Japan,
the FAIR/GSI in Germany, SPIRAL2/GANIL in France, and Facility for Rare Isotope
Beam (FRIB) in the USA \cite{Zhan06}.
These facilities offer
possibility to study the properties of nuclear matter or nuclei under the extreme conditions of large isospin asymmetries. Though at low incident energies, where fusion and related
phenomena are dominant, systematic  studies over isospin degree of freedom are available
\cite{Aror00},
no such studies are available at intermediate energies. One of the cause could be the
much more complex dynamics involved at intermediate incident energies.
Among various above mentioned phenomena nuclear stopping of the colliding matter has gained
a lot of interest since it gives us possibility to examine the degree of thermalization
or equilibration of the matter. \\
In a recent communication \cite{Liu01,Li05}, nuclear stopping
has been explored with reference to the isospin degree of freedom.
Nuclear stopping in heavy-ion collisions has been studied
by the means of rapidity distribution \cite{Khoa92} or by the asymmetry of
nucleonic momentum distribution
\cite{Baur88}.
Bauer and Bertsch \cite{Baur88} reported that, nuclear stopping is determined by both
the mean field and in-medium NN cross-sections.
Different authors suggested that the degree of approaching isospin equilibration provides a mean to prove the power of
nuclear stopping in HIC's \cite{Bass94}.\\
Interestingly, no systematical study is available in the literature on the
effect of isospin degree of freedom via symmetry energy on nuclear stopping.
This becomes much more important if one acknowledges that the effect of
symmetry energy could be altered in the presence of momentum dependent interactions,
which has become essential part of the interaction for any reasonable dynamical
model \cite{Khoa92,Bass94}. Thus, our aim is at least two folds:\\
\begin{itemize}
\item{We plan to understand the role of symmetry energy in the
presence of momentum dependent interactions in a systematic way.}\\
and
\item{To further examine how mass dependence alter the above findings. It is
worth mentioning that the study of the mass dependence is very essential for any meaningful
conclusion. }
\end{itemize}
This study is done
within the framework of isospin-dependent quantum molecular dynamics (IQMD) model,
which is discussed
in section-2. Section-3 contains the results and summary is presented in section-4.\\

%%%%%%%%%%%%%%%%%%%%%%%%%%%%%%%%%%%%%%%%%%%%%%%%%%%%%%%%%%%%%%%%%%%%%%%%%%%%%%%%%%%%%%%%%%%%%%%%%%%%%%%%%%%%%%% 
\section{ISOSPIN-DEPENDENT QUANTUM MOLECULAR DYNAMICS (IQMD) MODEL} 
The isospin-dependent quantum molecular dynamics (IQMD)\cite{Hart89} model
treats different charge states of
nucleons, deltas and pions
explicitly, as inherited from the VUU model \cite{Hart89}. The IQMD model has been used successfully
for the analysis of large number of observables from low to relativistic energies.
The isospin degree of
freedom enters into the calculations via symmetry potential, cross-sections and
nucleon-nucleon interactions.
The details about the elastic and inelastic cross-sections
for proton-proton and neutron-neutron collisions can be found in ref.\cite{Hart89}. \\
In this model,  baryons are represented by Gaussian-shaped density distributions
\begin{equation}
f_i(\vec{r},\vec{p},t) = \frac{1}{\pi^2\hbar^2}\cdot e^{-(\vec{r}-\vec{r_i}(t))^{2}\frac{1}{2L}}\cdot e^{-(\vec{p}-\vec{p_i}(t))^{2}\frac{2L}{\hbar^2}}.
\end{equation}
Nucleons are initialized in a sphere with radius $R= 1.12 A^{1/3}$ fm, in accordance with the liquid drop model. Each nucleon occupies a volume of $h^3$, so that phase space is uniformly filled. The initial momenta are randomly chosen between 0
and Fermi momentum($p_F$). The nucleons of target and projectile
interact via two and three-body Skyrme forces, Yukawa potential and momentum-dependent
interactions. These interactions are similar as used in the molecular dynamical models
like quantum molecular dynamics (QMD)\cite{Khoa92} and relativistic QMD
\cite{Lehm95}. The
isospin degree of freedom is treated explicitly by employing a symmetry potential and explicit Coulomb forces
between protons of colliding target and projectile. This helps in achieving correct distribution of protons and neutrons
within nucleus.\\
The hadrons propagate using Hamilton equations of motion:
\begin{equation}
\frac{d{r_i}}{dt}~=~\frac{d\it{\langle~H~\rangle}}{d{p_i}}~~;~~\frac{d{p_i}}{dt}~=~-\frac{d\it{\langle~H~\rangle}}{d{r_i}},
\end{equation}
with
\begin{eqnarray}
\langle~H~\rangle&=&\langle~T~\rangle+\langle~V~\rangle\nonumber\\
&=&\sum_{i}\frac{p_i^2}{2m_i}+
\sum_i \sum_{j > i}\int f_{i}(\vec{r},\vec{p},t)V^{\it ij}({\vec{r}^\prime,\vec{r}})\nonumber\\
& &\times f_j(\vec{r}^\prime,\vec{p}^\prime,t)d\vec{r}d\vec{r}^\prime d\vec{p}d\vec{p}^\prime .
\end{eqnarray}
 The baryon-baryon potential $V^{ij}$, in the above relation, reads as:
\begin{eqnarray}
V^{ij}(\vec{r}^\prime -\vec{r})&=&V^{ij}_{Skyrme}+V^{ij}_{Yukawa}+V^{ij}_{Coul}\nonumber\\
& &+V^{ij}_{mdi}+V^{ij}_{sym}\nonumber\\
&=&\left(t_{1}\delta(\vec{r}^\prime -\vec{r})+t_{2}\delta(\vec{r}^\prime -\vec{r})\rho^{\gamma-1}\left(\frac{\vec{r}^\prime +\vec{r}}{2}\right)\right)\nonumber\\
& & +~t_{3}\frac{exp(|\vec{r}^\prime-\vec{r}|/\mu)}{(|\vec{r}^\prime-\vec{r}|/\mu)}~+~\frac{Z_{i}Z_{j}e^{2}}{|\vec{r}^\prime -\vec{r}|}\nonumber\\
& & +t_{4}\ln^2[t_{5}(\vec{{p}_{i}}^\prime-\vec{p})^{2}+1]\delta(\vec{r}^\prime -\vec{r})\nonumber\\
& &+t_{6}\frac{1}{\varrho_0}T_3^{i}T_3^{j}\delta(\vec{r_i}^\prime -\vec{r_j}).
\label{s1}
\end{eqnarray}
Here $Z_i$ and $Z_j$ denote the charges of $i^{th}$ and $j^{th}$ baryon, and $T_3^i$, $T_3^j$ are their respective $T_3$
components (i.e. 1/2 for protons and -1/2 for neutrons). Meson potential consists of Coulomb interaction only.
The parameters $\mu$ and $t_1,.....,t_6$ are adjusted to the real part of the nucleonic optical potential. For the density
dependence of the nucleon optical potential, standard Skyrme-type parameterization is employed.
As is evident, we choose symmetry energy that depends linearly on the baryon
density. \\
The binary nucleon-nucleon collisions are included by employing collision term of well known VUU-BUU equation. The binary collisions
are done stochastically, in a similar way as are done in all transport models. During the propagation, two nucleons are
supposed to suffer a binary collision if the distance between their centroids
\begin{equation}
|r_i-r_j| \le \sqrt{\frac{\sigma_{tot}}{\pi}}, \sigma_{tot} = \sigma(\sqrt{s}, type),
\end{equation}
"type" denotes the ingoing collision partners (N-N, N-$\Delta$, N-$\pi$,..). In addition,
Pauli blocking (of the final
state) of baryons is taken into account by checking the phase space densities in the final states.
The final phase space fractions $P_1$ and $P_2$ which are already occupied by other nucleons are determined for each
of the scattering baryons. The collision is then blocked with probability
\begin{equation}
P_{block}~=~1-(1-P_1)(1-P_2).
\end{equation}
Delta decays are checked in an analogous fashion with respect to the phase space of the resulting nucleons. Recently, several studies have been devoted to pin down the strength of the
NN cross-section\cite{Khoa92}.\\
%%%%%%%%%%%%%%%%%%%%%%%%%%%%%%%%%%%%%%%%%%%%%%%%%%%%%%%%%%%%%%%%%%%%%%%%%%%%%%%%%%%%%%%%%%%%%%
\section{Results and Discussion}
The global stopping in heavy-ion collisions has been studied with the help of many different variables. In earlier studies, one used to relate the rapidity distribution with
global stopping. The rapidity distribution can be defined as \cite{Dhaw06,Wong94}:
\begin{equation}
Y(i)= \frac{1}{2}ln\frac{E(i)+p_{z}(i)}{E(i)-p_{z}(i)},
\end{equation}
where $E(i)$ and $p_z(i)$ are, respectively, the total energy and longitudinal momentum of $i^{th}$ particle. For a
complete stopping, one expects a single Gaussian shape. Obviously, narrow Gaussian 
indicate better thermalization
compared to broader Gaussian.\\
The second possibility to probe the degree of stopping is the anisotropy ratio (R) \cite{Liu01}:
\begin{equation}
R = \frac{2}{\pi}\frac{\left(\sum_{i}|p_{\perp}(i)|\right)}{\left(\sum_{i}|p_{\parallel}(i)|\right)},
\end{equation}
where, summation runs over all nucleons. The transverse and longitudinal momenta
are $p_{\perp}(i)$ =
$\sqrt{p_x^2(i) + p_y^2(i)}$ and $p_{\parallel}(i) = p_z(i)$, respectively.
Naturally, for a complete stopping,
$R$  should be close to unity.\\

\begin{figure}
\begin{center}
\includegraphics[width=0.75\textwidth]{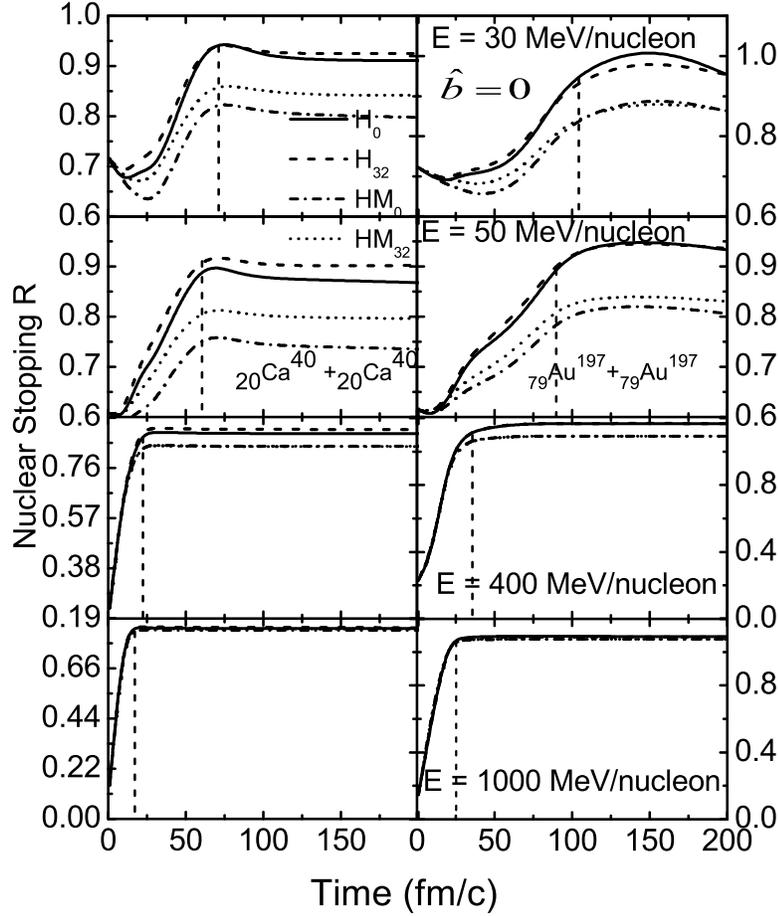}
\caption{The time evolution of anisotropy ratio $R$ for the reactions of $_{20}Ca^{40} + _{20}Ca^{40}$, and $_{79}Au^{197} + _{79}Au^{197}$. The panels from
top to bottom represent the scenario at beam energies 30, 50, 400 and 1000 MeV/nucleon,
respectively.}
\label{fig:2}
\end{center}
\end{figure}

\begin{figure}
\begin{center}
\includegraphics[width=0.75\textwidth]{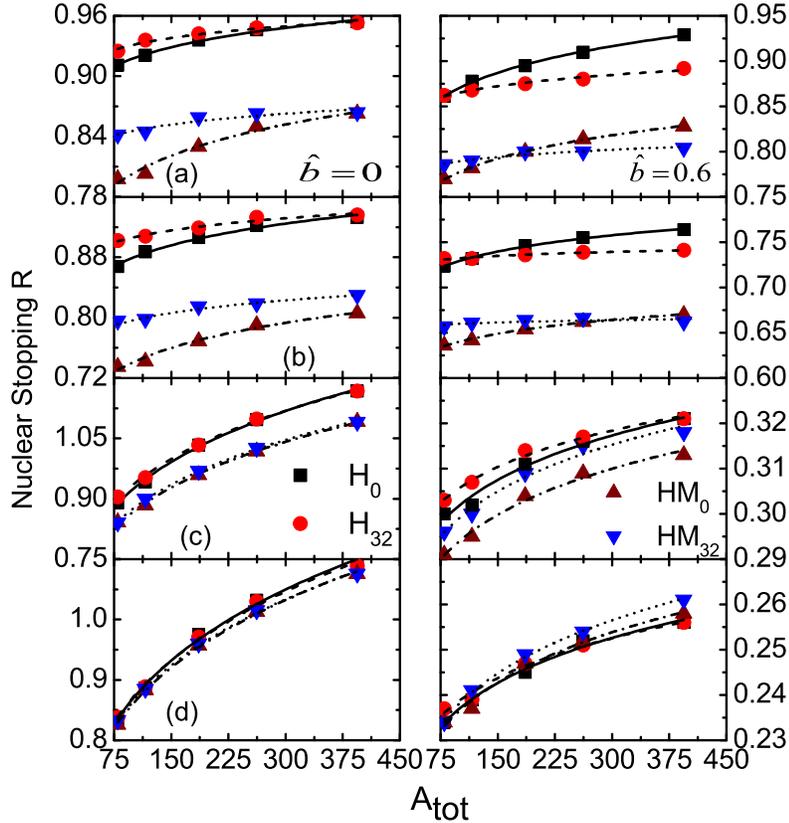}
\caption{ The final state anisotropy ratio $R$ as a function of
composite mass of the system $A_{tot}$ for
different possibilities of hard equation of state (discussed in text). The left and
right panels are at central and semi-peripheral geometries. The panels labeled
with a, b, c and d are at E= 30, 50, 400 and 1000 MeV/nucleon, respectively. All the curves are fitted with power law}
\label{fig:3}
\end{center}
\end{figure}

For the present study, simulations were carried out for the reactions $_{20}Ca^{40} + _{20}Ca^{40}$,
$_{28}Ni^{58} + _{28}Ni^{58}$, $_{41}Nb^{93} + _{41}Nb^{93}$, $_{54}Xe^{131} + _{54}Xe^{131}$ and$_{79}Au^{197} + _{79}Au^{197}$
at different beam energies ranging between 30 and 1000
MeV/nucleon at central and semi-peripheral geometries. The incident energy of
30 MeV/nucleon is the lowest limit for any semi-classical model.
Below this incident energy, quantum effects as well as Pauli blocking need to be
redefined.
A hard (H) and hard momentum dependent (HM) equation of state (EOS)
has been
employed with symmetry energy $E_{sym}$ = 0 and 32 MeV. The corresponding
values of the symmetry energy are indicated as subscript 
($H_0$, $H_{32}$, $HM_0$ and $HM_{32}$.).
It is worth mentioning that global stopping is insensitive toward the nature of static
equation of state. Our present attempt is to perform
a theoretical study with wider variation in the value of symmetry energy and to see
how much it can affect the heavy-ion dynamics. Though MDI destabilizes the nuclei, a careful
analysis is made by Puri {\it et al.} \cite{Sing01} and found that upto 200 fm/c,
emission of the nucleons with momentum dependent interactions is quite small.\\
\begin{figure}
\begin{center}
\includegraphics[width=0.75\textwidth]{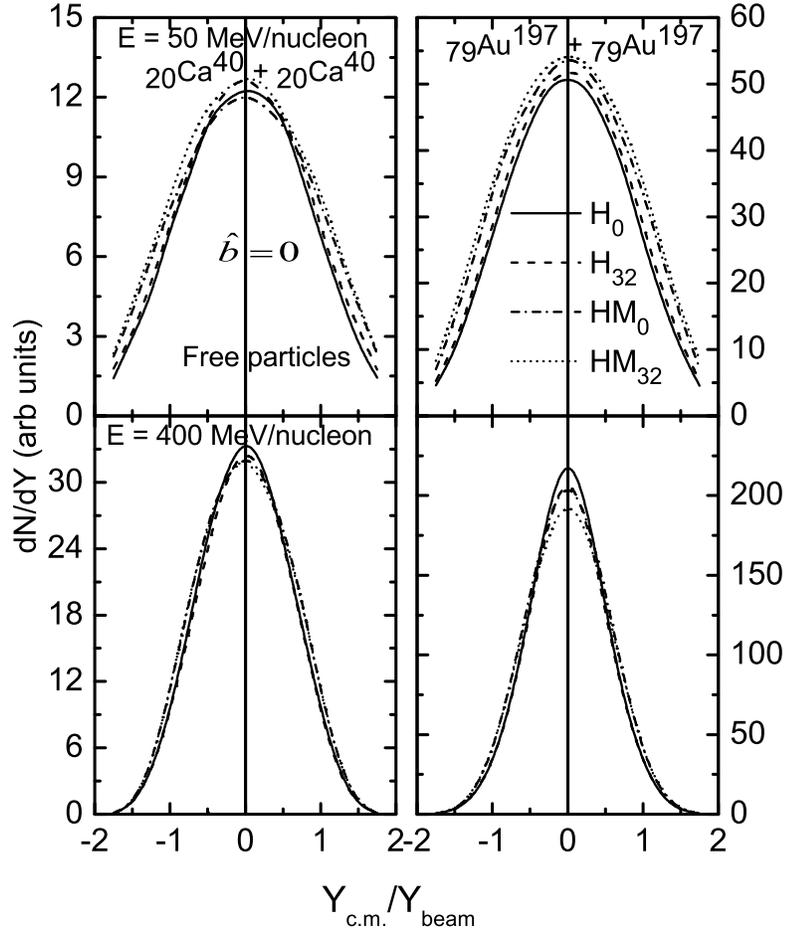}
\caption{ The rapidity distribution $\frac{dN}{dY}$ as a function of
reduced rapidity for free nucleons for the reactions of $_{20}Ca^{40} + _{20}Ca^{40}$, 
and $_{79}Au^{197} + _{79}Au^{197}$ for four different possibilities of hard 
equation of state. The top and bottom panels are at E = 50 and 
400 MeV/nucleon.}
\label{fig:1}
\end{center}
\end{figure}
Lets start with the aspect of rapidity distribution as a indicator for nuclear stopping.
As discussed earlier, nuclear stopping is a phenomena which originates from the participant zone.
To study the nuclear stopping in term of rapidity distribution in Fig.\ref{fig:1}, we 
display the rapidity distribution of free particles (which originates from the participant zone)
for different forms of the hard equation of state $(H_0$, $H_{32}$, $HM_0$ and $HM_{32})$.
We see that free particles emitted in the central collisions form a narrow Gaussian 
for the heavier system as well as at higher incident energy (say E = 400 MeV/nucleon). 
 It is evident from here that nuclear stopping is dominating with increase in size of the 
system as well as at higher incident energy (say E = 400 MeV/nucleon). 
With increase in incident energy, 
this behavior is not expected to be universal. It is clear from the Ref.\cite{Reis04} that
maximum stopping is observed around 400 MeV/nucleon. This study is done in detail 
in term of anisotropy ratio $R$ in the 
Figs. \ref{fig:2} and \ref{fig:3} of letter and found to be in supportive nature with 
Ref.\cite{Reis04}. 
On the other hand, nuclear stopping in term of rapidity distribution of protons is 
found to be weakly sensitive towards symmetry energy and momentum dependent interactions. 
This is due to the reason that there are not only free particles which originates from the 
participant zone. The other candidate which originates from this zone are the light 
charged particles (LCP's) $(2~\le~A~\le~4)$. In our recent communication\cite{Kuma10}, 
these LCP's are shown to be more sensitive towards symmetry energy due to pairing nature
and also good indicator for nuclear stopping.\\
For more meaningful, we will check the sensitivity of symmetry energy and momentum dependent
interactions on nuclear stopping in term of anisotropy ratio $R$, which is defined in 
detail in the above paragraph.
In Fig. \ref{fig:2}, we display the time evolution of the anisotropy ratio $R$ for the
central collisions of
$_{20}Ca^{40}~+~_{20}Ca^{40}$ (left panel) and $_{79}Au^{197}~+~_{79}Au^{197}$ (right panel).
The incident energies of 30,
50, 400 and 1000 MeV/nucleon are employed using the four different possibilities of
 hard equation of state
$(H_0$, $H_{32}$, $HM_0$ and $HM_{32})$.
Interestingly, the anisotropy ratio $R$, though, is insensitive towards
the symmetry energy, shows appreciable effect for the momentum dependent interactions.
At high enough incident energy, both effects (of momentum
dependent interactions as well of symmetry energy) wash away. For further test,
we simulated two reactions (i) $_{20}Ca^{40}+_{20}Ca^{40}$ and kept the same $N/Z$ ratio
by taking $_{100}X^{200}+_{100}X^{200}$ reaction. In other case (ii)
we took neutron rich reaction $_{79}Au^{197}+_{79}Au^{197}$ and simulated
the reactions of $_{16}X^{40}+_{16}X^{40}$ by keeping $N/Z$ ratio again same.
In both the cases,
the above trend holds good. Therefore, indicating that the above behavior is universal.
Even rapidity cuts to mid-rapidity region do not yield different results.
Below the beam energy of 50 MeV/nucleon, collision dynamics is governed
by the mean field. Therefore, interactions involving isospin particles like nn, np, pp
dominate the outcome and hence symmetry effects are visible.\\
As discussed earlier, $R$ approaches to
1 at low incident energies, indicating the isotropic nucleon momentum
distribution of the whole composite system. The behavior at 30 MeV/nuceon is
little different due to the fact that binary collisions do not play any role
and mean field will take larger time to thermalize the colliding nuclei. As beam energy
increases above the certain energy, $R$ starts decreasing from 1 towards the lower values,
indicating partial transparency.
This value of the beam energy, above which $R$ starts decreasing, depends on
the size of the system. In our observations for the reaction of
$_{79}Au^{197}~+~_{79}Au^{197}$, it is close to  400 MeV/nucleon.
This finding is similar to the one reported by W. Reisdorf {\it et al.} \cite{Reis04}.
The value of $R >1$, can be explained
by the preponderance of momentum perpendicular to beam direction \cite{Renf84}.
This is true for all equations of state. It is also seen that relaxation time
decreases with the increase in the beam energy, while,
increases with the increase in the mass of the colliding system. It shows that higher beam 
energies
and lighter systems
lead to more violent NN collisions and faster dissipation. This is consistent with the isospin equilibrium process
as shown by Li et.al. \cite{Li98}.\\
It will be further interesting to see whether the above findings have mass
dependence or not. This is particular important since the role of the momentum
dependent interactions and the symmetry energy depends on the size of the system.
For this, we display in Fig. \ref{fig:3}, the anisotropy ratio $R$ as a function of
the composite mass of the system ($A_{tot} = A_{T}+A_{P}$) at different
beam energies ranging from 30 to 1000 MeV/nucleon.
The left panel represents the results at central geometry, while, right panel is at semi-peripheral geometry.\\
Our findings are:
\begin{itemize}
\item{The anisotropy ratio $R$ increases with the composite mass of the system. This is
true for all
incident energies and impact parameters. This dependence becomes weaker as one 
moves from central
to semi-peripheral geometry.
It is due to the fact that nuclear stopping is governed by the participant zone only.
This is further supported by the fact that at higher incident
energies e.g. E = 1000 MeV/nucleon, the stopping is almost independent of the composite mass of the system at
semi-peripheral
geometry and almost 50$\%$  decrease is observed in the nuclear stopping as compared to central geometry.}\\
\item {The effect of the symmetry energy is visible below the Fermi energy. Same conclusion
was also reported in Fig. \ref{fig:2}.
The effect of the symmetry energy diminishes in the absence of 
momentum dependent interactions. Moreover, this
effect weakens at semi-peripheral geometries.}\\
and
\item {The importance of momentum dependent interactions is also visible in the
nuclear stopping.
This effect decreases as one moves from the low to higher incident energy and
from central to peripheral
geometry. At the higher incident energies e.g. at E = 1000 MeV/nucleon,
the anisotropy ratio is independent of
the equation of state and symmetry energy. It is worth mentioning that the inclusion of
momentum dependent interactions is found to suppress the binary collisions and as a
result is found to affect the sub threshold particle production as well as disappearance 
of collective flow \cite{Puri94,Khoa92}.}\\
\end{itemize}
 %%%%%%%%%%%%%%%%%%%%%%%%%%%%%%%%%%%%%%%%%%%%%%%%%%%%%%%%%%%%%%%%%%%%%%%%%%%%%%%%%%%%%%%%%%
\section{Summary}
Using the IQMD model, we have studied the nuclear stopping for isospin effects.
For this, we simulated the
reactions of
$_{20}Ca^{40} + _{20}Ca^{40}$, $_{28}Ni^{58} + _{28}Ni^{58}$, $_{41}Nb^{93} + _{41}Nb^{93}$,
$_{54}Xe^{131} + _{54}Xe^{131}$ and $_{79}Au^{197} + _{79}Au^{197}$ in the presence of
momentum dependent interactions and
symmetry energy.
The role of symmetry energy at low incident energy gets enhanced in the
presence of momentum dependent interactions. Further, we can conclude that maximum
stopping is obtained for the heavier systems at low incident energies in
central collisions in the absence of momentum dependent interactions implying
that momentum dependent interaction suppresses
the nuclear stopping.\\
%%%%%%%%%%%%%%%%%%%%%%%%%%%%%%%%%%%%%%%%%%%%%%%%%%%%%%%%%%%%%%%%%%%%%%%%%%%%%%%%%%%%%%%%%%%%%%%%%%%%%%%%%%%%%%%%%%
\ackn
This work is supported by the grant no. 03(1062)06/EMR-II, from the Council of Scientific
and Industrial Research (CSIR) New Delhi, govt. of India.

\section*{Refrences}  


\begin{thebibliography}{999}
\bibitem{Puri94} Puri R K {\it et al.} 1994 {\it Nucl. Phys. A} {\bf 575} 733; Puri R K
{\it et al.} 1996 {\it Phys. Rev. C} {\bf 54} R28; Puri R K {\it et al.} 2000
{\it J. Comp. Phys.} {\bf 162} 245; Kumar S, Kumar S and Puri R K 2010 
{\it Phys. Rev. C} {\bf 81}, 014611; Kumar S and Kumar S 2010 {\it Praman J. of Physics}
{\bf P-8343 (in press)}; Vermani Y K, Dhawan J K, Goyal S, Puri R K and Aichelin J 
2010 {\it J. of Phys. G: Nucl. and Part.} {\bf 37} 015105. 

\bibitem{Goss97} Gossiaux P B {\it et al.} 1997 {\it Nucl. Phys. A} {\bf 619} 379;
Kumar S {\it et al.} 1998 {\it Phys. Rev. C} {\bf 58} 3494; Fuchs C {\it et al.} 1996
{\it J. Phys. G: Nucl and Part.} {\bf 22} 131; Vermani Y K {\it et al.} 2009
{\it Eur. Phys. Lett.} {\bf 85} 62001.

\bibitem{Li98} Li B A {\it et al.} 1998 {\it Int. J. Mod. Phys. E} {\bf 7} 147;
Toro M D {\it et al.} 1999 {\it Prog. Nucl. Part. Phys.} {\bf 42} 125.

\bibitem{Zhan06} Zhan W {\it et al.} 2006 {\it Int. J. Mod. Phys. E} {\bf 15} 1941;
Yano Y  2007 "The RIKEN RI BEAM FACTORY PROJECT: A Status Report" {\it Nucl.
Inst. Meth. B} {\bf 261} 1009; http://www.gsi.de/fair/index\_e.html; http://www.ganinfo.
in2p3.fr/research/developments/spiral2; Whitepapers of the 2007 NSAC Long Range Plan Town Meeting, Jan., 2007, Chicago, http://dnp.aps.org.

\bibitem{Aror00} Arora R {\it et al.} 2000 {\it Eur. Phys. J. A} {\bf 8} 103; Puri R K
{\it et al.} 1991 {\it Phys. Rev. C} {\bf 43} 315; Puri R K {\it et al.} 1998
{\it Eur. Phys. J A} {\bf 3} 277; 1992 {\it Phys. Rev. C} {\bf 45} 1837; 1992 {\it J. Phys.
G: Nucl. and Part.} {\bf 18} 903; 1989 {\it Eur. Phys. Lett.} {\bf 9} 767;
Puri R K {\it et al.} 2005 {\it Eur. Phys. J. A} {\bf 23} 429; Malik S S {\it et al.}
1989 {\it Parmana J. Phys.} {\bf 32} 419; Dutt I and Puri R K 2010 {\it Phys. Rev. C} 
{\bf 81} 047601; {\it ibid.} {\bf 81} (in press) {\bf nuc-th/1004.0493}; {\it ibid.} 
{\bf 81} (in press); 
{\it ibid.} {\bf 81} (in press)

\bibitem{Liu01} Liu J Y {\it et al.} 2001 {\it Phys. Rev. Lett.} {\bf 86} 975;
Feng L Q and  Xia L Z 2002 {\it Chin. Phys. Lett.} {\bf 19} 321; Liu J Y, Guo W J,
Xing Y Z, Li X G and Gao Y Y 2004 {\it Phys. Rev. C} {\bf 70} 034610.

\bibitem{Li05} Li B A, Danielewicz P and Lynch W G 2005 {\it Phys. Rev. C} {\bf 71} 054603;
Luo X F {\it et al.} 2007 {\it ibid.} {\bf 76} 044902; Luo X F, Shao M, Dong X
and Li C, 2008 {\it Phys. Rev. C} {\bf 78} 031901.

\bibitem{Khoa92} Khoa D T {\it et al.} 1992 {\it Nucl. Phys. A} {\bf 619} 102;
Huang S W {\it et al.} 1993 {\it Phys. Lett. B} {\bf 298} 41; Batko G {\it et al.} 1994
{\it J. Phys. G: Nucl. and Part.} {\bf 20} 461; Singh J {\it et al.} 2000
{\it Phys. Rev. C} {\bf 62} 044617;
Sood A D {\it et al.} 2004 {\it Phys. Rev. C} {\bf 70} 034611; Kumar S {\it et al.} 1998
{\it Phys. Rev. C} {\bf 58} 1618, Huang S W {\it et al.} 1993 {\it Prog. Part. Nucl. Phys.}
{\bf 30} 105; Vermani Y K {\it et al.} 2009 {\it J. Phys. G: Nucl. and Part.}
{\bf 36} 105103;
Sood A D {\it et al.} 2004 {\it Phys. Lett. B} {\bf 594} 260; Lehmann E {\it et al.} 1996
{\it Z. Phys. A} {\bf 355} 55; Kumar S {\it et al.} 2008 {\it Phys. Rev. C } {\bf 78} 064602.

\bibitem{Baur88} Bauer W 1998 {\it Phys. Rev. Lett.} {\bf 61} 2534; Bertsch G F,
Brown G E, Koch V and Li B A 1998  {\it Nucl. Phys. A} {\bf 490} 745.

\bibitem{Bass94} Bass S A {\it et al.} 1994 {\it in GSI Annu. Report} {\bf p.66};
Li B A and Yennello S J 1995
{\it Phys. Rev.C} {\bf 52} R1746; Johnston H {\it et al.} 1996 {\it Phys. Lett. B}
{\bf 371} 186; Yennello S J {\it et al.} 1994 {\it Phys. Lett. B} {\bf 321} 15.

\bibitem{Hart89} Hartnack C {\it et al.} 1998 {\it Eur. Phys. J. A} {\bf 1} 151.

\bibitem{Lehm95} Lehmann E {\it et al.} 1995 {\it Phys. Rev. C} {\bf 51} 2113;
 1993 {\it Prog. Part. Nucl. Phys.} {\bf 30} 219.

\bibitem{Dhaw06} Dhawan J K, Dhiman N, Sood A D and Puri R K 2006 
{\it Phys. Rev. C} {\bf 74} 057901 and the
refrences within; Gossiaux P B and Aichelin J 1997 {\it Phys. Rev. C} {\bf 56} 2109.

\bibitem{Wong94} Wong C Y 1994 {\it Introduction to High-Energy Heavy-Ion Collisions} (World Scientific, Singapore).

\bibitem{Sing01} Singh J, Kumar S and Puri R K 2001 {\it Phys. Rev. C} {\bf 63} 054603;
Sood A D and Puri R K 2006
{\it Eur. Phys. J. A} {\bf 30} 571; Kumar S and Puri R K 1999 {\it Phys. Rev.
C} {\bf 60} 054607; Vermani Y K {\it et al.} 2009 {\it Phys. Rev. C} {\bf 79} 064613; 
Sood A K and Puri R K 2009 {\it Phys. Rev. C} {\bf 79} 064618.

\bibitem{Reis04} Reisdorf W {\it et al.} 2004 {\it Phys. Rev. Lett.} {\bf 92} 232301.

\bibitem{Kuma10} Kumar S., Kumar. S and Puri R. K. 2010 {\it Phys. Rev. C} {\bf 81} 014601. 

\bibitem{Renf84} Renforolt R E {\it et al.} 1984 {\it Phys. Rev. Lett.} {\bf 53} 763.

\end{thebibliography}
\end{document}